\documentclass[12pt]{iopart}

\usepackage{graphicx}
\begin{document}

\title[]{Directed tunneling of a prescribed number of dipolar bosons in shaken triple-well potentials}

\author{Xiaobing Luo$^{1,*}$, Yueming Wang$^{2}$, Xiaoguang Yu$^{1}$, Yu Guo$^{3}$, Guishu Chong $^{4,**}$, Donglan Wu$^{1}$, Qianglin Hu$^{1}$}

\address{$^{1}$Department of Physics, Jinggangshan University,
Ji'an 343009, China\\
 $^{2}$Department of Physics, Shanxi University, Taiyuan 030006, China\\
 $^{3}$School of Physics and Electronic Science, Changsha University of Science and Technology,
Changsha 410114, China\\
  $^{4}$Department of Physics, Hunan University, Changsha 410114, China\\
 $^*$ E-mail:xiaobingluo2013@aliyun.com\\
$^{**}$ E-mail:chonggs@hnu.edu.cn}

\begin{abstract}
We propose a scheme for precise control of tunneling dynamics of dipolar bosons in shaken triple-well potentials.
In the high-frequency regimes and under the resonance conditions, we have analytically and numerically demonstrated that we can transport a priori prescribed number of dipolar bosons along different pathways and different directions by adjusting the driving parameters. These results extend the previous many-body selective coherent destruction of tunneling (CDT) schemes for nondipolar bosons in double-well potentials[Phys. Rev. Lett.
103, 133002 (2009); Phys. Rev. A 86, 044102 (2012)], thus offering an efficient way to design the long-range coherent quantum transportation.
\end{abstract}

\maketitle

\section{Introduction}
~~~~In recent years, quantum-degenerate dipolar gases have attracted a great deal of
attention from both theoretical and experimental
studies\cite{Baranov}-\cite{Lahaye3}. In addition to the short-range and isotropic (s-wave) contact interaction, which is usually at work in
ultra-cold gases, long-range and anisotropic dipole-dipole
interaction (DDI) also plays a significant role in dipolar quantum gases
and gives rise to a rich variety of new physical properties.
A minimal system for direct visualization of the nonlocal characters of DDI is dipolar Bose-Einstein
condensate (BEC) in a triple-well potential, which is described by
a three-site Bose-Hubbard model with neighbor
interactions. This minimal system has been actively studied and known to display some novel features, such as
mesoscopic quantum superpositions\cite{Lahaye4}, interaction-induced coherence\cite{Xiong}, role of anisotropy\cite{Gallemi} and entanglement entropy\cite{Anna}.
In addition, dipolar Bose-Einstein condensates in triple-well potentials have been also investigated by ways of mean-field treatments\cite{Peter}-\cite{Zhang} and
multi-configuration time-dependent Hartree (MCTDH) method\cite{Chatterjee}.

Recently, there has been a burst of interest in experimental realization of quantum control with periodic
lattice shaking technique. Periodically shaking lattice leads to many important physics, such as coherent destruction of tunneling (CDT)\cite{Kierig,Lignier,Eckardt}, photon-assisted tunneling\cite{Sias}, Mott-insulator-superfluid transition\cite{Zenesini}, simulation of frustrated classical magnetism\cite{Struck1} and effective ferromagnetism\cite{Parker}, control of cotunneling and superexchange\cite{Bloch}, generation of synthetic gauge field\cite{Struck},  realization of Haldane
model\cite{Esslinger}, and so on. Among these intriguing aspects, CDT effect is a simple and powerful tool to control quantum tunneling dynamics\cite{Grossmann,Grifoni}, based on which two schemes for selective CDT of strongly interacting bosons in a symmetric double-well potential have been established by modulating the self-interaction strength\cite{Gong} or energy level unbalance\cite{Longhi}.
In these two many-body selective CDT schemes, the modulation
can be tuned in such a way that only an arbitrarily and a priori prescribed number of bosons are
allowed to tunnel from one well to the other. The sensitivity of CDT to particle number has also been demonstrated in an earlier work\cite{Creffield}, which enables the
self-trapped state to be used as a quantum beam splitter. As three-site system is a
paradigmatic model for longer array of lattice, extensive efforts have been paid to study the dynamics of periodically shaken
triple-well potentials\cite{Wang}-\cite{Luo2}. More recently, directed tunneling of $1$ or $N-1$ dipolar atoms with $N>2$ in a triple-well potential has been found numerically\cite{Lu2}. It naturally leads to another question: is it possible to
precisely control an arbitrarily prescribed number of dipolar bosons allowed to tunnel by shaking triple-well potential?

In this article, we suggest a method to control directed tunneling of a prescribed number of dipolar bosons
in periodically high-frequency shaken triple-well potentials, which extends the previous many-body selective CDT schemes for nondipolar bosons
in double-well potentials\cite{Gong,Longhi}. Under certain conditions, we analytically and numerically demonstrate that the driven three-site model can
be reduced to a double-well model, in which only tunneling between two adjacent wells is allowed.
We also reveal that direct tunneling of a priori prescribed number of dipolar bosons occurs along different pathways and different directions.
Our results may provide an additional possibility for
designing the long-range coherent quantum transportation.

\section{Model and high-frequency approximation}
~~~~The starting point of our analysis is provided by a driven
three-site Bose-Hubbard Hamiltonian in the presence of long-range DDI, which describes the
tunneling dynamics of dipolar bosons in a shaken triple-well
potential. Under the single-band tight-binding
approximation, by expanding the bosonic field operator
as $\hat{\psi}(\textbf{r})=\sum_{j=1}^{3} \phi_j(\textbf{r}) \hat{a}_{j}$ where $\phi_j(\textbf{r})$ is the Wannier state localized at the $j$th
site and $\hat{a}_{j}$ is the corresponding atom annihilation
operator, the three-site Bose-Hubbard Hamiltonian is given by\cite{Lahaye4,Lu2}
\begin{eqnarray}
\hat{H}&=&\hat{H}_{\rm{tun}}+\hat{H}_{\rm{int}}+\hat{H}_{\rm{ex}}\nonumber
\end{eqnarray}
with
\begin{eqnarray}
\hat{H}_{\rm{tun}}&=&-v\sum_{j=1}^{2}
(\hat{a}^{\dag}_{j+1}\hat{a}_{j}+\hat{a}^{\dag}_{j}\hat{a}_{j+1})
\nonumber\\
\hat{H}_{\rm{int}}&=&\frac{U_0}{2}\sum_{j=1}^{3}\hat{n}_{j}(\hat{n}_{j}-1)
+U_1(\hat{n}_{1}\hat{n}_{2}+\hat{n}_{2}\hat{n}_{3})+U_2\hat{n}_{1}\hat{n}_{3}
\nonumber\\
\hat{H}_{\rm{ex}}&=&\varepsilon(t)(\hat{a}^{\dag}_{1}\hat{a}_{1}-\hat{a}^{\dag}_{3}\hat{a}_{3}),\label{eq:H}
\end{eqnarray}
where $\hat{n}_j=\hat{a}^{\dag}_{j}\hat{a}_{j}$ is the particle number operator at the the $j$th
site, $v=\int d\textbf{r} \phi_j^*(\textbf{r})[-\nabla^2/2+V_{\rm{trap}}(\textbf{r})]\phi_{j+1}^*(\textbf{r})$
is the hopping (tunneling) rate between two sites, $U_0=g\int d\textbf{r}|\phi_j^*(\textbf{r})|^4+\int |\phi_j(\textbf{r})|^2\phi_j(\textbf{r}')|^2V_{\rm{dd}}(\textbf{r}-\textbf{r}')d\textbf{r}d\textbf{r}'$ characterizes the on-site interactions, $U_1=\int |\phi_j(\textbf{r})|^2\phi_{j+1}(\textbf{r}')|^2V_{\rm{dd}}(\textbf{r}-\textbf{r}')d\textbf{r}d\textbf{r}'$ is the coupling constant induced by nearest-neighbor DDI, and $U_2=\int |\phi_1(\textbf{r})|^2\phi_{3}(\textbf{r}')|^2V_{\rm{dd}}(\textbf{r}-\textbf{r}')d\textbf{r}d\textbf{r}'$ describes the next-nearest-neighbor interaction. In the above expressions, the natural unit $\hbar=m=1$ with $m$ being the atomic mass is used, $g = 4\pi a_s$ is the familiar short-range interaction constant with $a_s$ being the s-wave scattering length, $V_{\rm{trap}}(\textbf{r})$ is the time-independent triple-well scalar potential, and $V_{\rm{dd}}(\textbf{r}-\textbf{r}')$ is the DDI potential. Here the external driving field is applied in the form\cite{Sias,Lu2}, $\varepsilon(t)=\varepsilon_0+\varepsilon_1\cos(\omega t)$ with $\varepsilon_0$
and $\varepsilon_1$ being magnitudes of static and ac fields, $\omega$ the driving
frequency, respectively. The time-dependent part $\hat{H}_{\rm{ex}}$ of the Hamiltonian (\ref{eq:H}) can be experimentally realized by shaking (accelerating with a constant and modulating force) the triple-well potential\cite{Sias}.

The Hamiltonian (\ref{eq:H}) conserves the total number of particles.
We can expand the vector state $|\psi(t)\rangle$ of the
system in Fock space according to
\begin{eqnarray}
|\psi(t)\rangle=\sum\limits_{n_1,n_2,n_3}C_{n_1,n_2,n_3}|n_1,n_2,n_3\rangle,\label{state}
\end{eqnarray}
where $n_1+n_2+n_3=N$.
Substitution of the ansatz (\ref{state}) into the Schr\"{o}dinger equation
$i\partial_t|\psi(t)\rangle=\hat{H}|\psi(t)\rangle$,
one obtains the following coupled equations for the
probability amplitudes $C_{n_1,n_2,n_3}$

\begin{eqnarray}
i\dot{C}_{N,0,0}&=&V_{N,0,0}C_{N,0,0}+\kappa_0 C_{N-1,1,0}\nonumber\\
i\dot{C}_{N-n,n,0}&=&V_{N-n,n,0}C_{N-n,n,0}+\kappa_n C_{N-n-1,n+1,0}\nonumber\\&&+\kappa_{n-1} C_{N-n+1,n-1,0}+\nu_n C_{N-n,n-1,1}\nonumber\\&&(n=1,2,...,N-1)\nonumber\\
i\dot{C}_{0,N,0}&=&V_{0,N,0}C_{0,N,0}+\kappa_{N-1} C_{1,N-1,0}\nonumber\\&&+\kappa_{N-1} C_{0,N-1,1}
\nonumber\\
i\dot{C}_{0,n,N-n}&=&V_{0,n,N-n}C_{0,n,N-n}+\kappa_n C_{0,n+1,N-n-1}\nonumber\\&&+\kappa_{n-1} C_{0,n-1,N-n+1}+\nu_n C_{1,n-1,N-n}\nonumber\\&&(n=1,2,...,N-1)\nonumber\\
i\dot{C}_{0,0,N}&=&V_{0,0,N}C_{0,0,N}+\kappa_{0} C_{0,1,N-1}\nonumber\\
&&........,\nonumber\\
\label{coupled}
\end{eqnarray}
where we have set
\begin{eqnarray}
V_{n_1,n_2,n_3}&=&\langle n_1,n_2,n_3|(\hat{H}_{\rm{int}}+\hat{H}_{\rm{ex}})|n_1,n_2,n_3\rangle\nonumber\\
\kappa_n&=&-v\sqrt{(n+1)(N-n)},n=0,1,2...,N\nonumber\\
\nu_n&=&-v\sqrt{n},n=1,2...,N\nonumber\\
V_{N-n,n,0}
&=&F_{n}^1+\varepsilon_1\cos(\omega t)(N-n),n=0,1,2,...,N\nonumber\\
F_{n}^1&=&\frac{U_0}{2}[(N-n)(N-n-1)+n(n-1)]\nonumber\\&&+U_1(N-n)n+\varepsilon_0(N-n)
\nonumber\\
V_{0,n,N-n}
&=&F_{n}^2-\varepsilon_1\cos(\omega t)(N-n),n=0,1,2,...,N\nonumber\\
F_{n}^2&=&\frac{U_0}{2}[(N-n)(N-n-1)+n(n-1)]\nonumber\\&&+U_1(N-n)n
-\varepsilon_0(N-n)\nonumber\\
V_{N-n,n-1,1}
&=&F_{n}^3+\varepsilon_1\cos(\omega t)(N-n-1),n=1,2,...,N\nonumber\\
F_{n}^3&=&\frac{U_0}{2}[(N-n)(N-n-1)+(n-1)(n-2)]
\nonumber\\&&+U_1[(N-n)(n-1)+(n-1)]\nonumber\\&&+U_2(N-n)
+\varepsilon_0(N-n-1)\nonumber\\
V_{1,n-1,N-n}
&=&F_{n}^4-\varepsilon_1\cos(\omega t)(N-n-1),n=1,2,...,N\nonumber\\
F_{n}^4&=&\frac{U_0}{2}[(N-n)(N-n-1)+(n-1)(n-2)]
\nonumber\\&&+U_1[(N-n)(n-1)+(n-1)]\nonumber\\&&+U_2(N-n)
-\varepsilon_0(N-n-1).\nonumber\\
\label{coupled2}
\end{eqnarray}
Although it is difficult to obtain exact analytic solutions of
Eq.~(\ref{coupled}), we can approximately study some interesting
phenomena in the high-frequency regime with $\omega\gg v$. To that
end, we introduce the function transformation
\begin{eqnarray}
C_{n_1,n_2,n_3}(t)=B_{n_1,n_2,n_3}(t)\exp[-i\int_0^tV_{n_1,n_2,n_3}(t')dt'],\label{transformation}
\end{eqnarray}
Then, Eq.~(\ref{coupled}) is transformed to the coupled equations in
terms of new amplitudes $B_{n_1,n_2,n_3}(t)$
\begin{eqnarray}
i\dot{B}_{N,0,0}&=&\kappa_0 B_{N-1,1,0}\nonumber\\&&\times\exp[-i(F_1^1-F_0^1)t
+i\frac{\varepsilon_1}{\omega}\sin(\omega t)]\nonumber\\
i\dot{B}_{N-n,n,0}&=&\kappa_n B_{N-n-1,n+1,0}\nonumber\\&&\times\exp[-i(F_{n+1}^1-F_n^1)t+i\frac{\varepsilon_1}{\omega}\sin(\omega t)]\nonumber\\&&+\kappa_{n-1} B_{N-n+1,n-1,0}\nonumber\\&&\times\exp[-i(F_{n-1}^1-F_n^1)t-i\frac{\varepsilon_1}{\omega}\sin(\omega t)]\nonumber\\&&+\nu_n B_{N-n,n-1,1}\nonumber\\&&\times\exp[-i(F_{n}^3-F_n^1)t+i\frac{\varepsilon_1}{\omega}\sin(\omega t)]\nonumber\\&&(n=1,2,...,N-1)\nonumber\\
i\dot{B}_{0,N,0}&=&\kappa_{N-1} B_{1,N-1,0}\nonumber\\&&\times\exp[-i(F_{N-1}^1-F_N^1)t
-i\frac{\varepsilon_1}{\omega}\sin(\omega t)]\nonumber\\&&+\kappa_{N-1} B_{0,N-1,1}\nonumber\\&&\times\exp[-i(F_{N-1}^2-F_N^1)t
+i\frac{\varepsilon_1}{\omega}\sin(\omega t)]
\nonumber\\
i\dot{B}_{0,n,N-n}&=&\kappa_n B_{0,n+1,N-n-1}\nonumber\\&&\times\exp[-i(F_{n+1}^2-F_n^2)t
-i\frac{\varepsilon_1}{\omega}\sin(\omega t)]\nonumber\\&&+\kappa_{n-1} B_{0,n-1,N-n+1}\nonumber\\&&\times\exp[-i(F_{n-1}^2-F_n^2)t
+i\frac{\varepsilon_1}{\omega}\sin(\omega t)]\nonumber\\&&+\nu_n B_{1,n-1,N-n}\nonumber\\&&\times\exp[-i(F_{n}^4-F_n^2)t
-i\frac{\varepsilon_1}{\omega}\sin(\omega t)]\nonumber\\&&(n=1,2,...,N-1)\nonumber\\
i\dot{B}_{0,0,N}&=&\kappa_{0} B_{0,1,N-1}\nonumber\\&&\times\exp[-i(F_{1}^2-F_0^2)t
-i\frac{\varepsilon_1}{\omega}\sin(\omega t)]\nonumber\\
&&........\nonumber\\
\label{coupled3}
\end{eqnarray}
To produce the directed motion of dipolar bosons, we assume that the following resonance conditions are satisfied
\begin{eqnarray}
(U_0-U_1)=(U_1-U_2)=\omega,\label{resonance1}\\\varepsilon_0=m\omega, m=0,\pm 1,\pm 2,...\label{resonance2}
\end{eqnarray}
Under such resonant conditions, the system will exchange energy with
the driving field to bridge the energy gap resulting from strong
interactions and static tilt, and the energy scale of the system
becomes characterized by the natural tunneling coefficient $v$. When
the shaking frequency is much larger than the energy scale $v$, the
amount of change in $B_{n_{1}, n_{2}, n_{3}}(t)$ during a period,
 $T=2\pi/\omega$, can be regarded as being infinitesimal. Thus Eq. (\ref{coupled3})
can be integrated approximately over a period  $2\pi/\omega$ by
supposing that $B_{n_{1}, n_{2}, n_{3}}(t)$ are constants. By
averaging the rapidly oscillating exponential terms in
Eq.~(\ref{coupled3}), we obtain a group of approximate equations for
the evolution of the amplitudes $B_{n_1,n_2,n_3}(t)$
\begin{eqnarray}
i\dot{B}_{N,0,0}&=&\kappa_0 B_{N-1,1,0}\mathcal{J}_{-(N-1)-m}(\frac{\varepsilon_1}{\omega})\nonumber\\
i\dot{B}_{N-n,n,0}&=&\kappa_n B_{N-n-1,n+1,0}\mathcal{J}_{-(N-2n-1)-m}(\frac{\varepsilon_1}{\omega})\nonumber\\&&+\kappa_{n-1} B_{N-n+1,n-1,0}\mathcal{J}_{-(N-2n+1)-m}(\frac{\varepsilon_1}{\omega})\nonumber\\&&+\nu_n B_{N-n,n-1,1}\mathcal{J}_{-(N-1)-m}(\frac{\varepsilon_1}{\omega})\nonumber\\&&(n=1,2,...,N-1)\nonumber\\
i\dot{B}_{0,N,0}&=&\kappa_{N-1} B_{1,N-1,0}\mathcal{J}_{(N-1)-m}(\frac{\varepsilon_1}{\omega})\nonumber\\&&+\kappa_{N-1} B_{0,N-1,1}\mathcal{J}_{-(N-1)-m}(\frac{\varepsilon_1}{\omega})
\nonumber\\
i\dot{B}_{0,n,N-n}&=&\kappa_n B_{0,n+1,N-n-1}\mathcal{J}_{(N-2n-1)-m}(\frac{\varepsilon_1}{\omega})\nonumber\\&&+\kappa_{n-1} B_{0,n-1,N-n+1}\mathcal{J}_{(N-2n+1)-m}(\frac{\varepsilon_1}{\omega})\nonumber\\&&+\nu_n B_{1,n-1,N-n}\mathcal{J}_{(N-1)-m}(\frac{\varepsilon_1}{\omega})\nonumber\\&&(n=1,2,...,N-1)\nonumber\\
i\dot{B}_{0,0,N}&=&\kappa_{0} B_{0,1,N-1}\mathcal{J}_{(N-1)-m}(\frac{\varepsilon_1}{\omega})\nonumber\\
&&........,\nonumber\\
\label{coupled4}
\end{eqnarray}
where $\mathcal{J}_n$ is the $n$th-order Bessel function of first kind. Equation (\ref{coupled4})
is effective in description of the tunneling dynamics of the original system for resonance driven
case, which is the basis of the following analysis.

\section{Directed tunneling of dipolar bosons under preestablished condition}
\label{sec:tunneling} ~~~~ Generally, it is hard to solve
analytically the large numbers of coupled equations in
(\ref{coupled4}),  as the dimension of the Hilbert space increases
sharply with $N$. In what follows, we are interested in the special
case $\mathcal{J}_{-(N-1)-m}(\frac{\varepsilon_1}{\omega})=0$ or
$\mathcal{J}_{(N-1)-m}(\frac{\varepsilon_1}{\omega})=0$, where the
coupling among the equations in (\ref{coupled4}) is partly removed
so that some of the equations become closed and analytical. We now
proceed to illustrate how to control the tunneling processes of a
precisely defined number of bosons.

When $\mathcal{J}_{-(N-1)-m}(\frac{\varepsilon_1}{\omega})=0$ is
selected, the system dynamics is limited in a subspace spanned by
states $|N-1,1,0\rangle, |N-2,2,0\rangle,...,|N-n,n,0\rangle,...,
|0,N,0\rangle$, and the motion of equation (\ref{coupled4}) becomes
\begin{eqnarray}
i\dot{B}_{N-n,n,0}&=&\kappa_n B_{N-n-1,n+1,0}\mathcal{J}_{-(N-2n-1)-m}(\frac{\varepsilon_1}{\omega})\nonumber\\&&+\kappa_{n-1} B_{N-n+1,n-1,0}\mathcal{J}_{-(N-2n+1)-m}(\frac{\varepsilon_1}{\omega})\nonumber\\&&(n=1,2,...,N-1)\nonumber\\
i\dot{B}_{0,N,0}&=&\kappa_{N-1} B_{1,N-1,0}\mathcal{J}_{(N-1)-m}(\frac{\varepsilon_1}{\omega}).
\nonumber\\
\label{coupled5}
\end{eqnarray}
It can be learned from Eq.~(\ref{coupled5}) that the tunneling
pathway between wells 2 and 3 is shut off and only tunneling between
wells 1 and 2 is allowed.

On the other hand, applying
$\mathcal{J}_{(N-1)-m}(\frac{\varepsilon_1}{\omega})=0$ to
Eq.~(\ref{coupled4}) yields
\begin{eqnarray}
i\dot{B}_{0,N,0}&=&\kappa_{N-1} B_{0,N-1,1}\mathcal{J}_{-(N-1)-m}(\frac{\varepsilon_1}{\omega})
\nonumber\\
i\dot{B}_{0,n,N-n}&=&\kappa_n B_{0,n+1,N-n-1}\mathcal{J}_{(N-2n-1)-m}(\frac{\varepsilon_1}{\omega})\nonumber\\&&+\kappa_{n-1} B_{0,n-1,N-n+1}\mathcal{J}_{(N-2n+1)-m}(\frac{\varepsilon_1}{\omega})\nonumber\\&&(n=1,2,...,N-1).\nonumber\\
\label{coupled6}
\end{eqnarray}
In this case, the tunneling between wells 1 and 2 is prohibited and
the tunneling passage between wells 2 and 3 is switched on. So far,
it has been shown, under the selective CDT conditions
$\mathcal{J}_{-(N-1)-m}(\frac{\varepsilon_1}{\omega})=0$ or
$\mathcal{J}_{(N-1)-m}(\frac{\varepsilon_1}{\omega})=0$, the driven
three-site model can be reduced to an effective double-well model
(\ref{coupled5}) or (\ref{coupled6}) respectively, in which only
tunneling between two adjacent wells is allowed. Next, we will
present the underlying physics behind this decoupling.

 Suppose there is a Fock state $|N-i,i,0\rangle$, describing that
 $N$ bosons occupy the left and central wells with the right well
 empty. As this Fock state
 changes from $|N-i,i,0\rangle$ to $|N-i,i-1,1\rangle$ (hence one particle is released to the right
 well), there is a corresponding loss of energy
\begin{eqnarray}
\Delta
 E_1=[N(U_1-U_2)-(U_0-U_1)]+\varepsilon_0.
\label{Loss}
\end{eqnarray}
Here, the energy loss $\Delta E_1$ comes from two parts: one is the
loss
 of interaction energy, the other is the energy difference due to
 the static tilt of triple-well potential. When the frequency of
 driving is chosen such that
\begin{eqnarray}
k\omega=\Delta E_1,\label{match}
\end{eqnarray}
where $k$ is an integer, the energy of $k$ photons bridges the
energy gap between states $|N-i,i,0\rangle$ and $|N-i,i-1,1\rangle$,
and, as a result, the tunneling contact disabled by both static tilt
and strong interaction can be restored in general. This is analogous
to photon-assisted tunneling. The energy match of Eq.~(\ref{match})
is realized under the resonance conditions
 (\ref{resonance1})-(\ref{resonance2}). In the fast modulation $\omega\gg v$ regime,
 the effective tunneling coefficient between resonant states $|N-i,i,0\rangle$ and $|N-i,i-1,1\rangle$
 is approximately renormalized by $\mathcal{J}_k(\varepsilon_1/\omega)$. However, at particular values of the amplitude of driving
 field,
$\mathcal{J}_k(\varepsilon_1/\omega)=\mathcal{J}_{(N-1)+m}(\frac{\varepsilon_1}{\omega})=\mathcal{J}_{-(N-1)-m}(\frac{\varepsilon_1}{\omega})=0$,
CDT will occur and the tunneling dynamics to the right well will be
frozen.

Under such circumstance, the system dynamics is limited in the left
and central wells, which is governed by the effective motion of
equation (\ref{coupled5}). From Eq.~(\ref{coupled5}) we know that
the effective tunneling coefficient between states $|N-i,i,0\rangle$
and $|N-i+1,i-1,0\rangle$ is rescaled by a factor of
$\mathcal{J}_{-(N-2i+1)-m}(\frac{\varepsilon_1}{\omega})$. This
renormalization of tunneling coefficient can also be interpreted as
 multiphoton resonances between states $|N-i,i,0\rangle$ and
$|N-i+1,i-1,0\rangle$. As such, if the additional condition
\begin{eqnarray}
\mathcal{J}_{-(N-2i+1)- m}(\frac{\varepsilon_1}{\omega})=0
 \label{condition_1}
\end{eqnarray}
is satisfied, no more particle is allowed to tunnel from the central
to the left well (hence the transition
$|N-i,i,0\rangle\rightarrow|N-i+1,i-1,0\rangle$ becomes prohibited)
since the central well has already released $N-i$ particles to the
left well. Therefore, a desired and prescribed number $N-i$ of
particles can be allowed to tunnel from the central to the left well
provided that the CDT condition
$\mathcal{J}_{-(N-1)-m}(\frac{\varepsilon_1}{\omega})=0$ and the
condition of Eq.~(\ref{condition_1}) are simultaneously satisfied.
Such two conditions can be achieved by setting
\begin{eqnarray}
\mathcal{J}_{i-1}(\frac{\varepsilon_1}{\omega})=0,
\varepsilon_0=-(N-i)\omega.
\end{eqnarray}

Up to now, we have established the conditions for controlling a
definite number of  bosons allowed to tunneling from the central to
the left well from an energetics argument. The main point of such
selective control of tunneling processes is that \emph{at
multiphoton resonances, the system undergoes simultaneously two
kinds of CDT effects (one is the decoupling of the left-center wells
from the right, and the other is particle-number-dependent CDT in
the left-center wells).}

In the meantime, the decoupling (right-center from left)
 and the particle-number-dependent CDT in the right-center
 wells can be achieved in a similar manner. The particle number dependence of the tunneling
coefficient between the central and left wells is different from
that of the tunneling coefficient between the central and right
wells. This discrepancy is due to the exactly opposite values of
loss of energy resulting from constant tilt between the transitions
of $|N-i,i,0\rangle\rightarrow|N-i+1,i-1,0\rangle$ and
$|0,i,N-i\rangle\rightarrow|0,i-1,N-i+1\rangle$, in spite of the
same interaction energy loss. Thus, we can switch the directed
tunneling of a precisely defined number of bosons along the
center-left pathway to the center-right pathway, by only reversing
the constant tilt $\varepsilon_0$. \emph{The fact that static tilt
breaks the inversion symmetry in the triple well allows for a
selective control of tunneling processes along different pathways
and different directions.}

Following the analysis mentioned above, we summarize the main
results as follows:

(i) The direct tunneling of a definite number $N-i$ of particles
from the central well to the left or to the right can be realized
under the preestablished conditions
\begin{eqnarray}
\mathcal{J}_{i-1}(\frac{\varepsilon_1}{\omega})=0,
\varepsilon_0=\mp(N-i)\omega.
\end{eqnarray}
Here $``-"$ is for the directed tunneling along the pathway from the
central to the left well, and $``+"$  for the pathway from the
central to the right well.

(ii) For the initial states $|N-1,1,0\rangle$ ($|0,1,N-1\rangle$),
which may be prepared through many-body state engineering using
measurements and fixed unitary dynamics\cite{Pedersen}, the
conditions for controlling a desired number $i-1$ bosons allowed to
tunneling from the left (right) to the central well is given by
\begin{eqnarray}
\mathcal{J}_{i}(\frac{\varepsilon_1}{\omega})=0,
\varepsilon_0=\mp(N-i-1)\omega,
 \label{condition_2}
\end{eqnarray}
where $``-"$ in the second equality corresponds to the direct
tunneling from the left to the central well, $``+"$ to the direct
tunneling from the right to the central well.

(iii) Additionally, if all the Bessel functions in
Eq.~(\ref{coupled5}) or (\ref{coupled6}) are non-vanishing, $N-1$
bosons participate in the tunneling process from the central well to
the left or right. Thus we can give alternative conditions for such
two tunneling processes as
$\mathcal{J}_{-(N-1)-m}(\frac{\varepsilon_1}{\omega})=0,
\varepsilon_0=m\omega\neq -(N-i)\omega, (i\neq 1)$ and
$\mathcal{J}_{(N-1)-m}(\frac{\varepsilon_1}{\omega})=0,
\varepsilon_0=m\omega\neq (N-i)\omega, (i\neq 1)$, respectively.

\section{Numerical experiments}
~~~~We have checked our theoretical predictions by direct numerical
simulations of Eq.~(\ref{eq:H}). In all simulations, we typically
assumed $N=4$ bosons and $U_0=75, U_1=40, U_2=5, \omega=35, v=1$, for which the resonance condition (\ref{resonance1}) is satisfied.
\begin{figure*}[htp]
\center
\includegraphics[width=5.8cm]{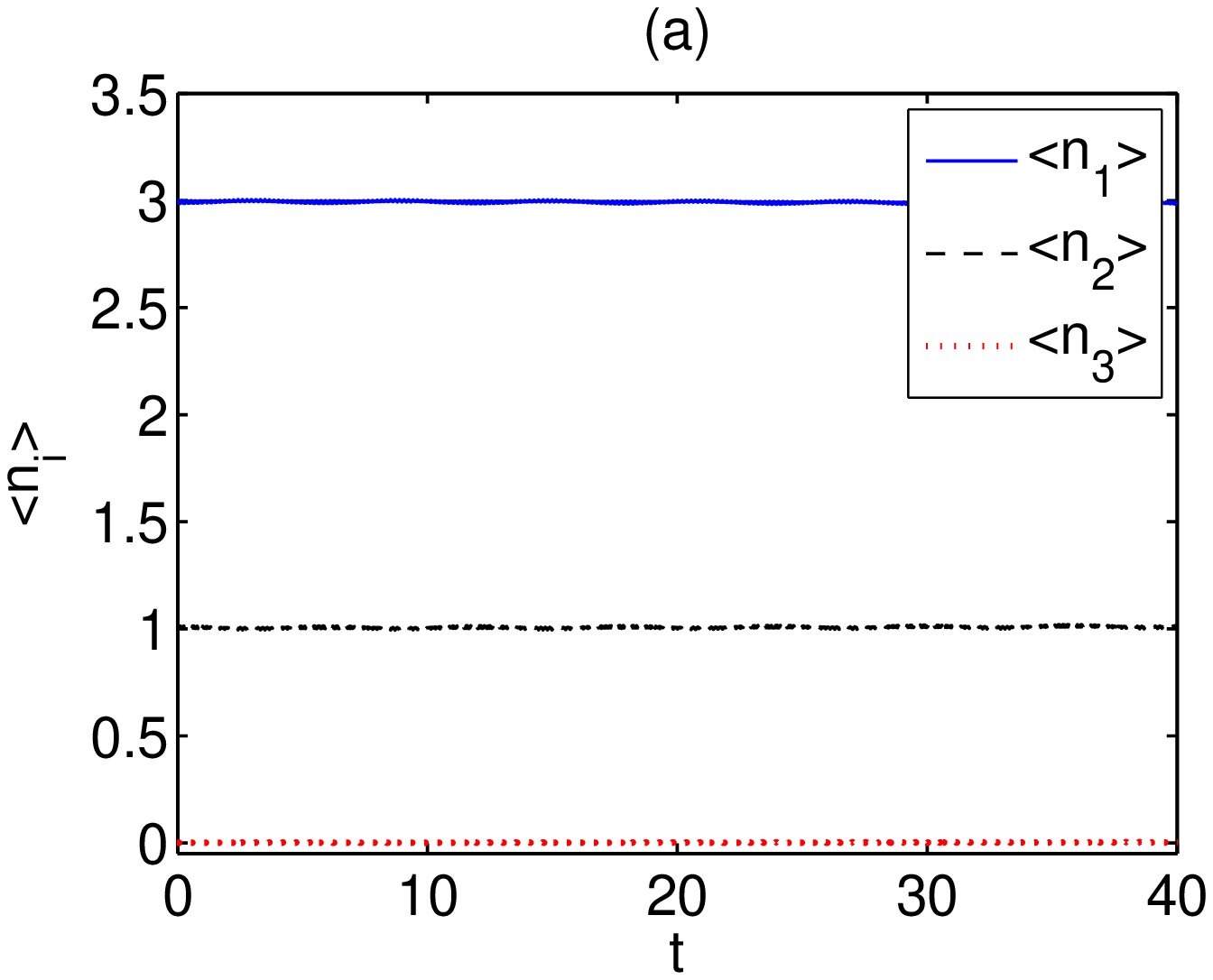}
\includegraphics[width=5.8cm]{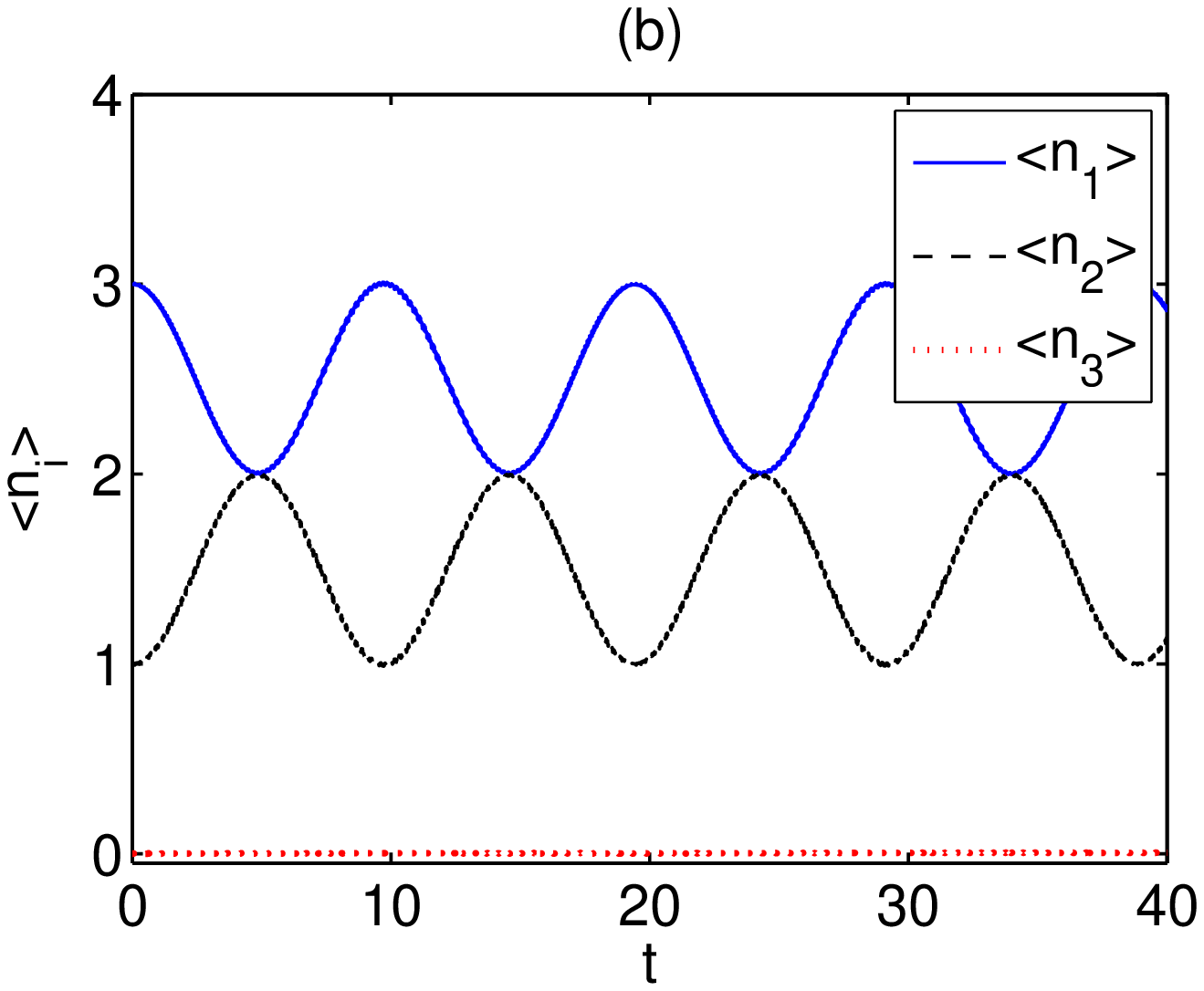}
\includegraphics[width=5.8cm]{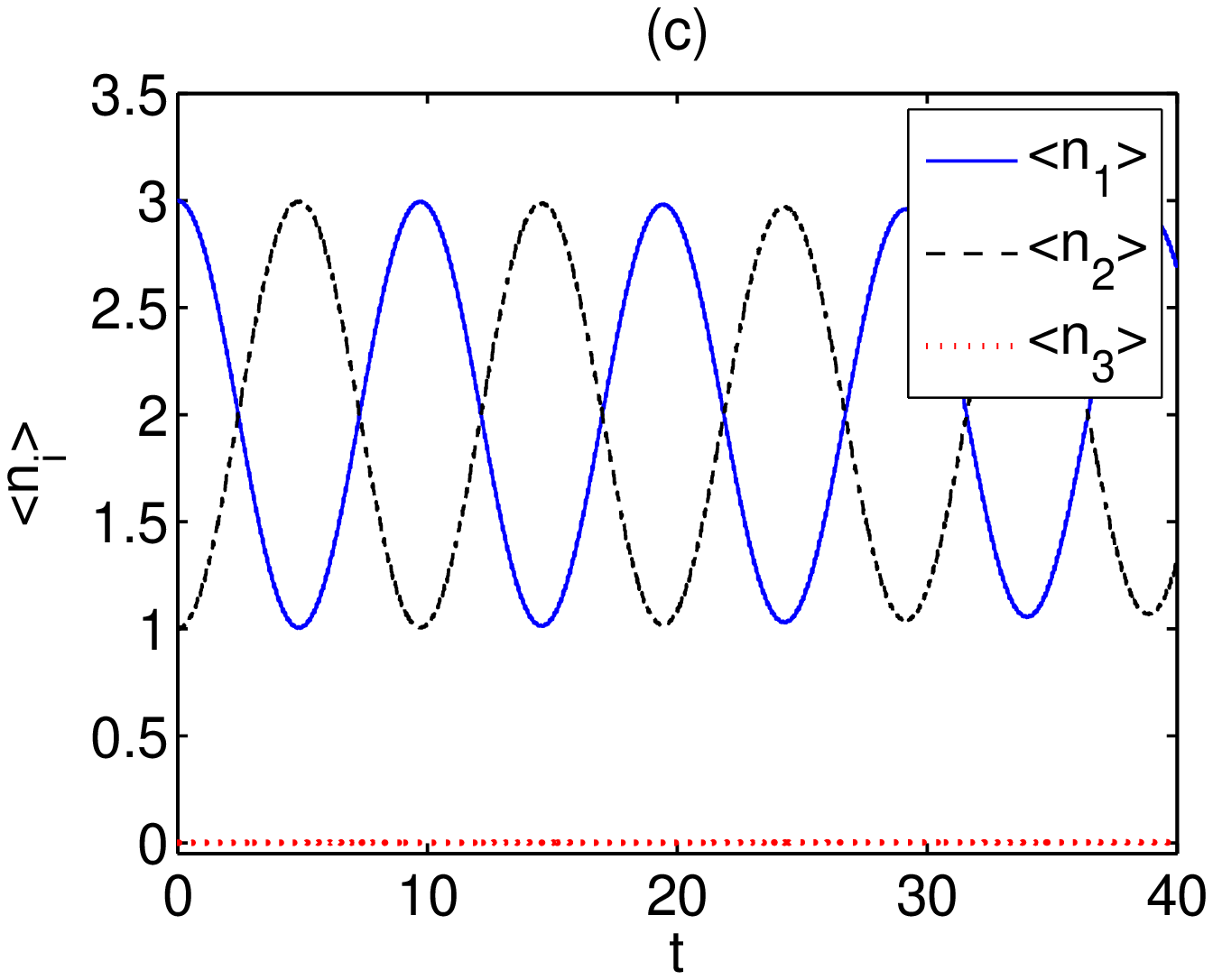}
\caption{(color online) Averages $\langle n_i(t)\rangle,i=1,2,3$ for different driving parameters: (a) $\varepsilon_0=-2\omega$, $\varepsilon_1/\omega=3.8317$; (b) $\varepsilon_0=-\omega$, $\varepsilon_1/\omega=5.1356$; (c) $\varepsilon_0=0$, $\varepsilon_1/\omega=6.3802$. The initial state is $|3,1,0\rangle$ and the other
parameters are chosen as $U_0=75, U_1=40, U_2=5, \omega=35, v=1$. The three cases of Figs.~\ref{fig1}(a)-(c) correspond to the directed tunneling of zero, one, and two bosons from the left to central well, respectively.}\label{fig1}
\end{figure*}

\begin{figure*}[htp]
\center
\includegraphics[width=5.8cm]{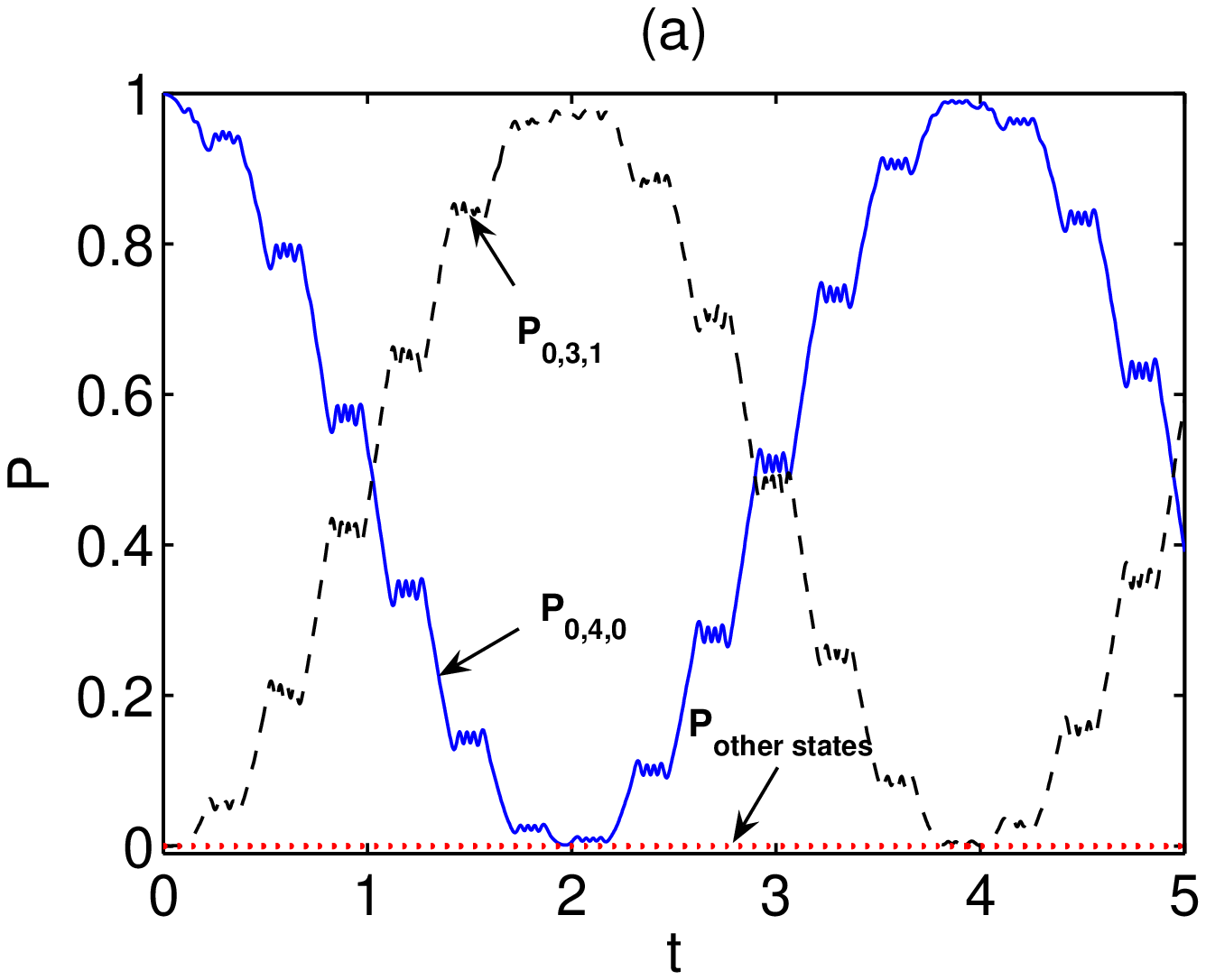}
\includegraphics[width=5.8cm]{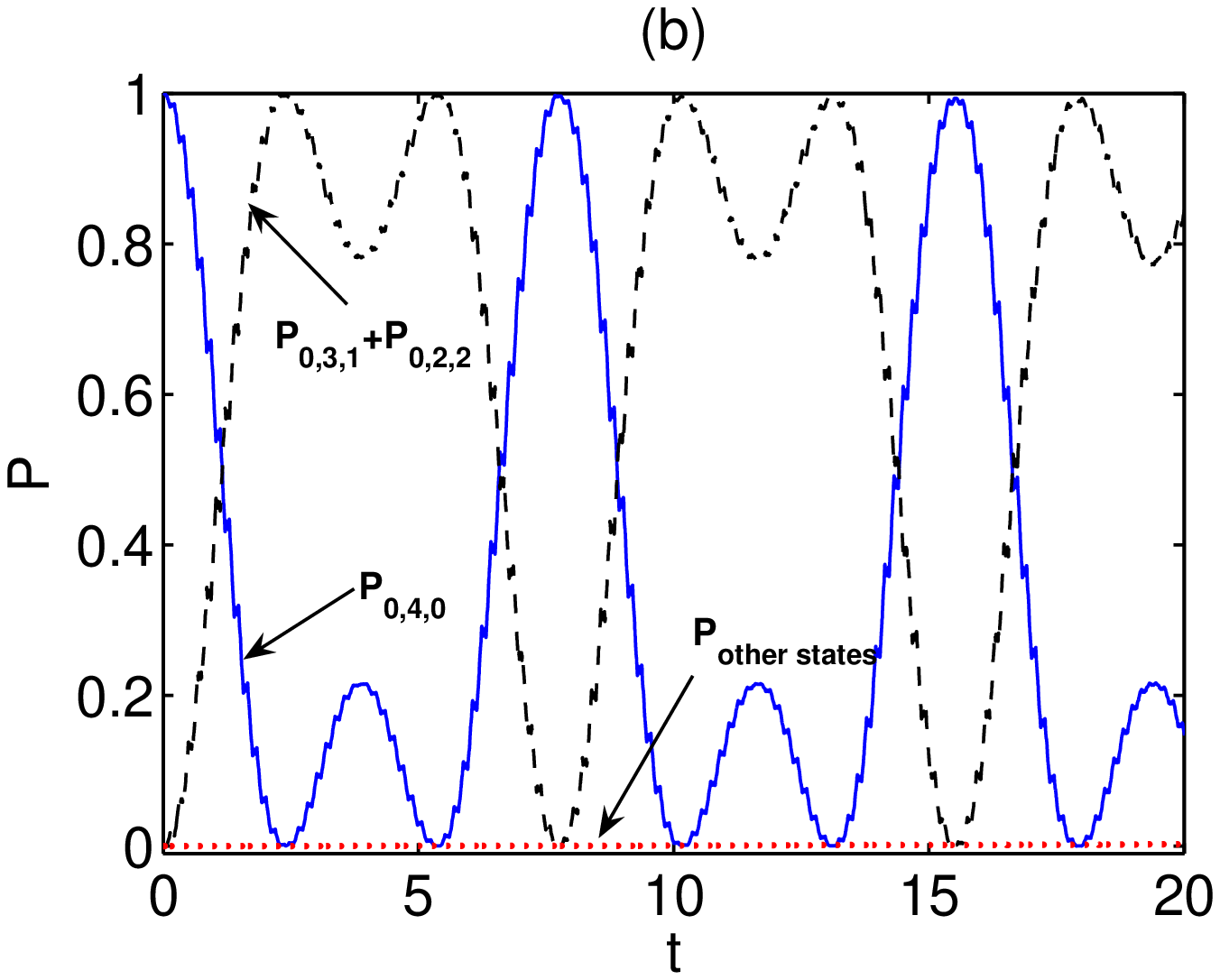}
\includegraphics[width=5.8cm]{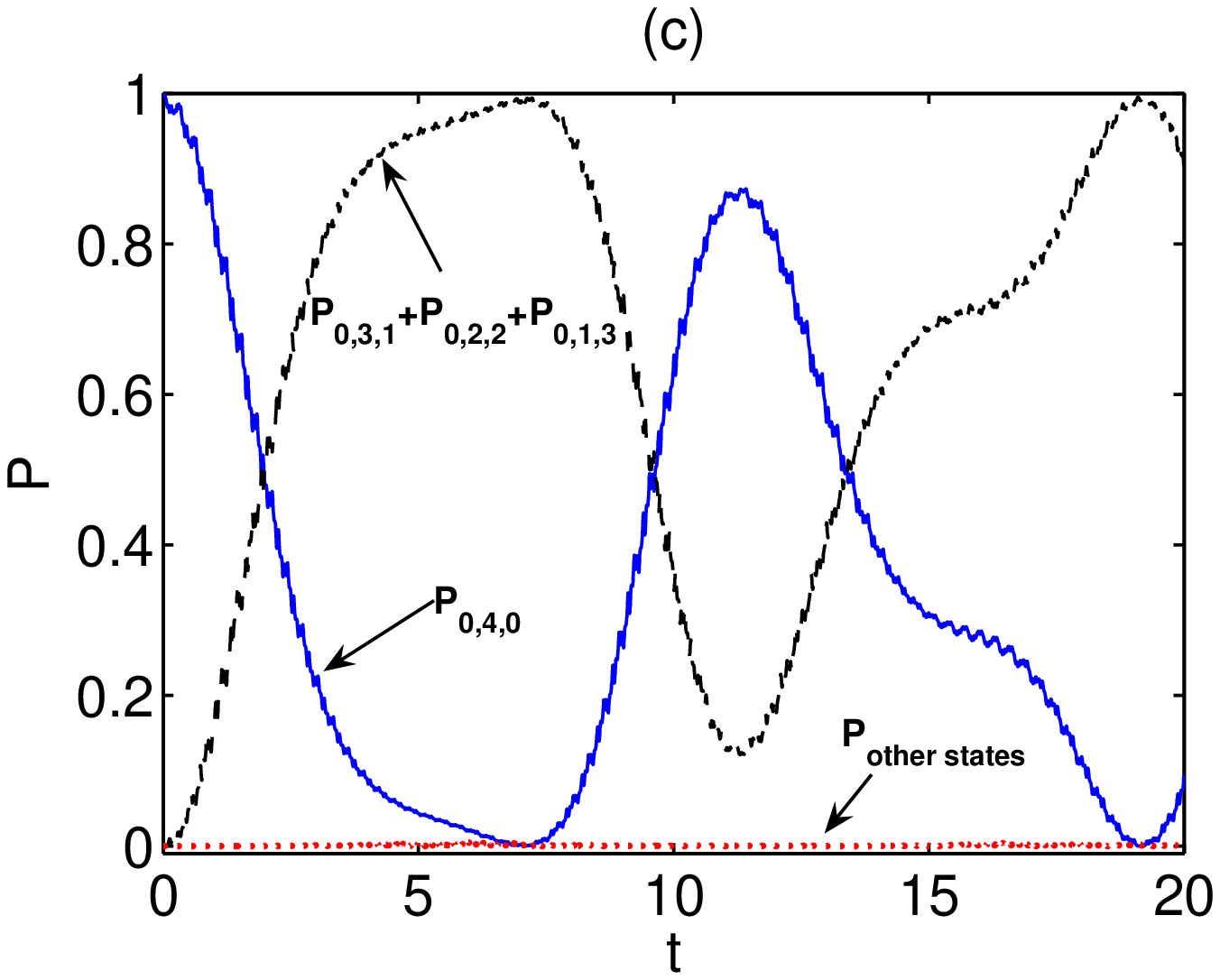}
\caption{(color online) Time evolution of the probability
distributions for system (\ref{eq:H}) for different driving
parameters: (a) $\varepsilon_0=\omega$,
$\varepsilon_1/\omega=5.1356$; (b) $\varepsilon_0=2\omega$,
$\varepsilon_1/\omega=7.0156$; (c) $\varepsilon_0=-\omega$,
$\varepsilon_1/\omega=7.5883$. $N=4$ bosons are initially prepared
in the central well. The other parameters are $U_0=75, U_1=40,
U_2=5, \omega=35, v=1$. The three cases of Figs.~\ref{fig2}(a)-(c)
represent $|0,4,0\rangle\rightarrow |0,3,1\rangle$,
$|0,4,0\rangle\rightarrow
C_{0,3,1}|0,3,1\rangle+C_{0,2,2}|0,2,2\rangle$, and
$|0,4,0\rangle\rightarrow
C_{0,3,1}|0,3,1\rangle+C_{0,2,2}|0,2,2\rangle+C_{0,1,3}|0,1,3\rangle$,
respectively.} \label{fig2}
\end{figure*}

As an example, Figs.~\ref{fig1}(a)-(c) show the evolution of
$\langle n_i \rangle=\langle \psi(t)| \hat{a}^{\dag}_{i}\hat{a}_{i}
|\psi(t)\rangle$, numerically computed from Eq.~(\ref{eq:H}) for the
initial state $|3,1,0\rangle$ and for three different values of
driving parameters satisfying conditions (\ref{condition_2})(therein
the second condition having $``-"$ sign has been applied). In the
first case for $\varepsilon_0=-2\omega$
[$\varepsilon_0=-(N-i-1)\omega$, $i=1$] and
$\varepsilon_1/\omega=3.8317$
[$\mathcal{J}_{i}(\frac{\varepsilon_1}{\omega})=\mathcal{J}_{1}(\frac{\varepsilon_1}{\omega})=0$],
all $\langle n_i \rangle$ maintain their initial values,
demonstrating that no particle is allowed to tunnel. In the second
case for $\varepsilon_0=-\omega$ [$\varepsilon_0=-(N-i-1)\omega$,
$i=2$] and $\varepsilon_1/\omega=5.1356$
[$\mathcal{J}_{i}(\frac{\varepsilon_1}{\omega})=\mathcal{J}_{2}(\frac{\varepsilon_1}{\omega})=0$],
$\langle n_1 \rangle$ oscillates between 3.0 and 2.0, and $\langle
n_3 \rangle$ remains negligible at all times, which demonstrate that
one particle is allowed to tunnel from the left to right well.
Similarly, in the third case for $\varepsilon_0=0$
[$\varepsilon_0=-(N-i-1)\omega$, $i=3$] and
$\varepsilon_1/\omega=6.3802$
[$\mathcal{J}_{i}(\frac{\varepsilon_1}{\omega})=\mathcal{J}_{3}(\frac{\varepsilon_1}{\omega})=0$],
we see that $\langle n_1 \rangle$ oscillates between 3.0 and 1.0,
$\langle n_2\rangle$ oscillates between 1.0 and 3.0, and  $\langle
n_3\rangle$ is still zero, which show that two particles are allowed
to tunnel from the left to right well. These numerical results
verify firmly the occurrence of direct tunneling of dipolar bosons
in a prescribed number under preestablished conditions.

As demonstrated in Section (\ref{sec:tunneling}), for all particles
initially occupying the central well, the tunneling of $N-i$ bosons
from the central to right well occurs when the relations
$\mathcal{J}_{i-1}(\frac{\varepsilon_1}{\omega})=0$ and
$\varepsilon_0=(N-i)\omega$ are satisfied, and the tunneling of
$N-i$ bosons occurs from the central to left well when the condition
$\mathcal{J}_{i-1}(\frac{\varepsilon_1}{\omega})=0$ still holds and
the constant tilt $\varepsilon_0$ is switched to $-(N-i)\omega$. To
confirm these predictions, we calculate numerically time evolution
of the probability distribution $P_{n_1,n_2,n_3}=|C_{n_1,n_2,n_3}|^2
$ for the system (\ref{eq:H}) with the initial state
$|0,4,0\rangle$. Three sets of our results are shown in
Figs.~\ref{fig2}(a)-(c). In the first set for $\varepsilon_0=\omega$
[$\varepsilon_0=(N-i)\omega$, $i=3$] and
$\varepsilon_1/\omega=5.1356$
[$\mathcal{J}_{i-1}(\frac{\varepsilon_1}{\omega})=\mathcal{J}_{2}(\frac{\varepsilon_1}{\omega})=0$],
we see that full transition between states $|0,4,0\rangle$ and
$|0,3,1\rangle$ occurs, without tunneling to other states. This
means that only one of $N$ dipolar bosons is allowed to be
transferred from well 2 to well 3. In the second set for
$\varepsilon_0=2\omega$ [$\varepsilon_0=(N-i)\omega$, $i=2$] and
$\varepsilon_1/\omega=7.0156$
[$\mathcal{J}_{i-1}(\frac{\varepsilon_1}{\omega})=\mathcal{J}_{1}(\frac{\varepsilon_1}{\omega})=0$],
we see that only transition between $|0,4,0\rangle$ and
$C_{0,3,1}|0,3,1\rangle+C_{0,2,2}|0,2,2\rangle$ are allowed, in
which two bosons participate in the tunneling process along the path
between wells 2 and 3.  In the third set for $\varepsilon_0=-\omega$
[$\varepsilon_0\neq (N-i)\omega$, $N-i=0,1,2$] and
$\varepsilon_1/\omega=7.5883$
[$\mathcal{J}_{N-1-m}(\frac{\varepsilon_1}{\omega})=\mathcal{J}_{4}(\frac{\varepsilon_1}{\omega})=0$],
it can be seen that the system experiences transition between state
$|0,4,0\rangle$ and superposition state
$\sum_{n=1}^{N-1}C_{0,N-n,n}|0,N-n,n\rangle$ with zero population at
other states, in which only the pathway between wells 2 and 3 is
switched on and $N-1$ bosons participate in the tunneling process
along this tunneling path.

Moreover, we have demonstrated some other situations by direct
numerical simulations of Eq.~(\ref{eq:H}). In Ref.~\cite{Lu2}, the
authors have numerically exhibited the directed tunneling of one
particle from the central to right well and directed tunneling of
$N-1$ particles from the central to left well, which can be viewed
as two explicit examples in our work. Our primary purpose here is to
provide a method for precise control of the tunneling of a priori
prescribed number of dipolar bosons along different pathways and
along different directions, which will greatly facilitate the
control of quantum states. The numerical results (not shown)
demonstrate that our theoretical predictions are still applicable
even when the interactions and the driving frequency are not very
large compared to tunneling rate, indicating that our proposal is
more realistic than what it seems.

\begin{figure*}[htp]
\center
\includegraphics[width=8cm]{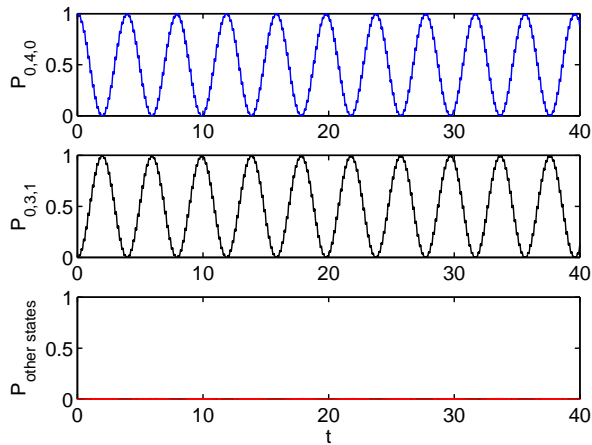}
\caption{(color online) Evolution of the probability distributions
versus time for a moderate value of detuning $\beta=5$. The
parameters are the same as those in Fig.~\ref{fig2}(a) except for
next-nearest-neighbor interaction $U_2=10 (\beta=5)$.} \label{fig3}
\end{figure*}

Before concluding, we present some remarks on our theoretical
predictions. Dissipation like particle loss presents a major
obstacle for long-time coherent control of quantum states. The
lifetime of bosonic system is principally limited by dissipative
three-body interactions, which reduces rapidly with the decrease of
particle number. Like in the schemes of Refs.~\cite{Gong} and
\cite{Longhi}, high-frequency approximation is valid when driving
frequency $\omega$ is of the same order of magnitude or larger than
the coupling constants $\kappa_n,\nu_n$, which requires relatively
small numbers of bosons $N$. Presently, moderately small systems of
2-10 atoms can be precisely prepared with high
purity\cite{Weitenberg,Winkler,Folling} and long lifetime up to a
few tenths of a second\cite{Campbell}, indicating that coherent
directed tunneling processes should be observable in our considered
system. Our proposal requires that three interaction parameters
(on-site $U_0$, nearest-neighbor $U_1$, and next-nearest neighbor
$U_2$) are subjected to the resonance condition (\ref{resonance1}).
According to the theoretical calculation in Ref.~\cite{Lahaye4}, the
ratio of the nearest-neighbor to next-nearest-neighbor interaction
$U_1/U_2$ depends on the geometry of potential and varies from 4 to
8. Note that on-site interaction $U_0$ results from short-range
interaction and DDI, in which contact interaction between particles
can be very precisely controlled by means of
 Feshbach resonance\cite{Cornish}, and DDI may be manipulated by varying the
shape of a dipolar BEC, the dipole polarization axis, and the
trapping geometry\cite{Baranov}-\cite{Lahaye3}. We have numerically
simulated our main findings by extending the rigorous resonance
condition (\ref{resonance1}) to nonresonance case
$U_0-U_1=U_1-U_2+\beta$. The numerical results show that our
theoretical predictions are relatively tolerant against moderate
changes in detuning $\beta$. As an example, Fig.~\ref{fig3} shows
that tunneling process of Fig.~\ref{fig2}(a) is insensitive to
detuning $\beta$. The results imply it is easier to realize
experimentally these directed selective-tunneling effects. With the
constantly advancing lattice shaking
techniques\cite{Kierig}-\cite{Esslinger}, we expect our results can
be tested in the realistic experimental setups.

\section{Conclusions}
~~~~In summary, we have theoretically studied a generalization of many-body selective CDT, which enables one to control a priori prescribed number of dipolar bosons allowed to tunnel in shaken triple-well potentials. In the high-frequency regimes and under the resonance conditions, through rotating-wave (or
high-frequency averaging) approximation method, we obtain a group of effectively coupled equations for the evolution of probability amplitudes. By adjusting the driving parameters, we can decouple these coupled equations and thus establish the conditions for directed tunneling of a priori prescribed number of dipolar bosons. Under the preestablished conditions, we can transport a desired number of dipolar bosons along different pathways and different directions.

We expect that our findings can be extended to other multi-well systems and thus give us a deep insight into the tunneling dynamics of dipolar condensates in optical lattices.
 Our results may provide an opportunity to manipulate the tunneling of an array of dipolar bosons and may be useful for efficient quantum information processing and atomic device designing.



\section*{Acknowledgments}
~~~~X. Luo and Y. Wang thank Congjun Wu for his providing us with an opportunity of visiting Department of Physics at
University of California, San Diego, where part of this work is carried out.
The work was supported by the NSF of China under Grants 11465009, 11165009, 10965001, 10904035, the Program for New Century Excellent Talents in University of
Ministry of Education of China (NCET-13-0836), Atomic and Molecular
Physics Key Discipline of Jiangxi Province, the financial support from China Scholarship Council, and Scientific and
Technological Research Fund of Jiangxi Provincial Education
Department under Grant No. GJJ14566. Y. Guo was supported by the NSF of China (Grant No. 11105020) and the Scientific Research Fund of Hunan Provincial Education Department under Grant No. 13B134.

\end{document}